\documentclass{article}

\usepackage{arxiv}

\usepackage[utf8]{inputenc} 
\usepackage[T1]{fontenc}    
\usepackage{url}            
\usepackage{booktabs}       
\usepackage{amsfonts}       
\usepackage{nicefrac}       
\usepackage{microtype}      
\usepackage{lipsum}

\usepackage{amsmath}        
\usepackage{amssymb}        
\usepackage{graphicx}       
\usepackage{bm}             
\usepackage{float}          
\usepackage{hyperref}       

\usepackage[nomarkers,figuresonly]{endfloat} 
\usepackage{subfig}         

\title{Route Identification in the National Football League}

\author{
  Dani ~Chu\thanks{\url{https://danichusfu.github.io/}} \\
  Department of Statistics \& Actuarial Sciences\\
  Simon Fraser University \\
  Burnaby, BC V5A 1S6 \\
  \texttt{dani\_chu@sfu.ca} \\
   \And
 Matthew ~Reyers\thanks{\url{https://matt-reyers.netlify.com/}} \\
  Department of Statistics \& Actuarial Sciences\\
  Simon Fraser University \\
  Burnaby, BC V5A 1S6 \\
  \texttt{matthew\_reyers@sfu.ca} \\
  \And
 James ~Thomson \\
  Department of Statistics \& Actuarial Sciences\\
  Simon Fraser University \\
  Burnaby, BC V5A 1S6 \\
  \texttt{james\_thomson\_2@sfu.ca} \\
  \And
 Lucas ~Wu\\
  Department of Statistics \& Actuarial Sciences\\
  Simon Fraser University \\
  Burnaby, BC V5A 1S6 \\
  \texttt{yifan\_wu@sfu.ca} \\
}

\begin{document}
\maketitle

\begin{abstract}

Tracking data in the NFL is a sequence of spatial-temporal measurements that vary in length depending on the duration of the play. In this paper, we demonstrate how model-based curve clustering of observed player trajectories can be used to identify the routes run by eligible receivers on offensive passing plays. We use a Bernstein polynomial basis function to represent cluster centers, and the Expectation Maximization algorithm to learn the route labels for each of the 34,698 routes run on the 6,963 passing plays in the data set. We go on to suggest ideas for new potential receiver metrics that account for receiver deployment. The resulting route labels can also be paired with film to enable streamlined queries of game film.
\end{abstract}



\section{Introduction}

Curve clustering is the process of finding a latent class structure for observations of functional data and has wide applications across industries such as biology, finance and environmental science \cite{aghabozorgi}. Many methodologies have been developed to deal with such data \cite{jiguo}. These methodologies fall into four main methods; shape-based, compression based dissimilarity, feature based, and model-based clustering \cite{aghabozorgi}.

This paper will focus on a model-based route clustering methodology using Gaussian mixture methods. These have been used in previous literature to cluster gene expression \cite{mcnicholas}, recognize Arabic characters \cite{alshaher}, and distinguish regions based on temperature data \cite{Bouveyron2011}. In sport literature, model-based clustering has been used to cluster swimming progression in competition \cite{Leroy_2018}, and in basketball to cluster player trajectories in the National Basketball Association (NBA) \cite{Miller2017Possessions}. 

The player trajectory data used by Miller and Bornn has been available to NBA teams since 2013 \cite{nba}. Similar data has recently been made available in the NFL by Next-Gen Stats and affiliate organizations \cite{nfl_tracking}. The player tracking data in the NFL is collected differently than in the NBA, utilizing a chip in the shoulder pads of players for the entirety of collection rather than computer vision tools. Although the data is collected in different ways, the player tracking data is fundamentally the same and allows for analysts in each sport to identify player movement over time. Capturing these insights has proven valuable in the NBA - leading the NFL to pursue the same approach.

For this reason, the NFL hosted the inaugural NFL Big Data Bowl, a competition that released player tracking data to the public for the first time, in January of 2019 to help teams gain insights from this data. Previous work in football had relied on play-by-play event data which was popularized and made readily available by the nflscrapr package \cite{R-nflscrapR} for the \textsf{R} Project for Statistical Computing. This led to work such as Expected Points Added, Win Probability Added, and Wins Above Replacement models in football \cite{nflwar} which had already been readily available in sports such as golf \cite{broadie}, basketball \cite{stern} and baseball \cite{open_war_b}.

The data used in the NFL Big Data Bowl was NFL Next-Gen Stats tracking data which was gathered by Zebra Technologies and Wilson Sporting Goods through the use of radio-frequency identification (RFID) chips. These chips measure the field position (x, y) of each player and the ball at ten equally spaced points per second. Key events are listed at the moment they occur, such as a snap, a tackle, or a fumble. Using this information, the shape of each route run by every receiver can be considered a finite function, with a known start and ending point.

\subsection{Routes in the NFL}
A route in football is the path or pattern that an eligible receiver runs on a passing play. Coaches and quarterbacks plan the route for every eligible receiver prior to the start of the play. In some cases receivers may have an option to run one of many predetermined routes where they decide on the specific route to run based on how the defense is set up. These are called option routes. Additionally, the predetermined routes are subject to change. A quarterback can communicate to his receivers new routes before the play begins or a receiver can make an adjustment while running the route.

Football teams of all levels dedicate staff to tagging videos with the routes run by receivers on passing plays. Doing so is long and tedious work but it allows teams to query plays that meet certain search criteria. For example, to prepare for a playoff game, a coach may want video of all plays where a specific team has receivers run a three route combination including a flat, in, and post route. The same coach may want to evaluate his own team's success in all plays involving receivers on the same side of the field that run an out route and a go route. Further, a defensive back in preparation for a marquee match-up with a top wide receiver may want to watch film of all the plays where that wide receiver ran a post route.

With the new player tracking data and the long amount of hours spent tagging film, the automated detection and labeling of routes is of interest to football teams. It is of specific interest to not just detect and label these routes, but to be able to identify the known patterns that exist within the NFL. Although not restricted as a set, a relatively well defined route tree exists in which many realized routes are variations of with the addition of in-play noise. Capturing this structure within the data would lend validity to the utilized method and provide a direct use case for the NFL and its affiliate organizations. Before the availability of this player tracking data, attempts were made to use computer vision and machine learning to identify routes and formations from game film \cite{ajmeri}. Now that the NFL offers tracking data, machine learning techniques have been instead focused on route identification through feature based supervised learning techniques \cite{hochstedler}.

We propose using a model-based unsupervised learning approach to clustering routes using the new player tracking data. We will implement this by using  B\'{e}zier curves to define cluster means and then learn the cluster parameters and membership probabilities using the Expectation-Maximization algorithm. We will then label the clusters and provide direction for the use of these labels in further analysis. 

Labelling the routes run on a given play provides the information needed for a more nuanced analysis of receiver play in the NFL. Statistics like targets over expectation and air yards over expectation require information about receiver deployment on passing plays. Automating this labelling will help save hundreds of manual labour hours tagging plays and make querying plays of interest easier.

This sharing of ideas and methodologies across sports and industries leads us to present our route identification methods for routes in the NFL.

\section{Bernstein Basis \& B\'{e}zier Curves}

In pursuing a model-based approach to clustering routes, we will make specific use of B\'{e}zier curves. First established by Pierre \'{E}tienne B\'{e}zier, these curves are capable of representing complicated "free-form" shapes of infinite points. The fundamentals of this approach originate in Bernstein basis polynomials \cite{bernstein}. The connection between Bernstein Basis Polynomials and B\'{e}zier curves is straightforward. First, we can define the basis polynomials for a degree $P$ on $t \in [0, 1]$ by 
\begin{equation}
    b_p^P(t) =  \binom{P}{p} t^p (1 - t)^{P - p},\ p = 0, \ldots, P 
\end{equation}
Extending these polynomials to the B\'{e}zier setting is then done through applying a weight to each term of the polynomial, called control points, by
 \begin{equation} \label{eq:bezier}
         \mathbf{B}(t ; \bm{\theta}) = \sum_{p = 0}^P \theta_p b_p^P (t),\ t \in [0, 1],
 \end{equation}
with control points $\bm{\theta}_0, \ldots, \bm{\theta}_P$. The output for a given input $t$, given the control points $\bm{\theta}$, are the coordinates of the corresponding point on the B\'{e}zier curve. The collection of outputs over the inputs $t$ then forms the B\'{e}zier curve.

B\'{e}zier curves are parametric curves and are easy to implement. They generalize well to higher dimensions and make for an excellent general tool. Though their application has been extensively explored in computer graphics, groundwork has been started with player tracking data in the NBA \cite{Miller2017Possessions}. It is worth noting that a desirable property of these B\'{e}zier curves is that they naturally handle the comparing of routes that differ in duration. We will rely on this property frequently due to the variability in duration of NFL routes. Our work will look to expand this introductory work into the NFL for receiver routes.

\section{Fitting B\'{e}zier Curves}

Adapting the work of \cite{gaffney}, let $\mathbf{Y}$ be a set  $n$ observed player trajectories, $\{\mathbf{y}_1, \ldots, \mathbf{y}_n \}$. Each observed curve $y_i$ has $m_i$ observations along its trajectory. These points describe the location at each observed time in two dimensional space. Therefore, $y_i$ is a $m_i \times 2$ matrix. 

Each observed trajectory is measured at times $\mathbf{t}_i$. We assume that each curve can be described by a B\'{e}zier curve with degree $P$, defined by $\bm{\theta}$, which is a $(P + 1 ) \times 2$ matrix, and an additive Gaussian error term $\bm{\epsilon}_i$, a $m_i \times 2$ matrix. The $j$th term of $\bm{\epsilon}_i$ is $\epsilon_{ij} \sim \mathcal{N}(\mathbf{0}, \sigma^2 \mathbf{I})$, and  $\epsilon_{ij}, \sigma^2$ which are both $1 \times 2$ matrices ($i=1,2,\dots,n$, $j=1,2,\dots,m_{i}$).

Since each trajectory can be described by a B\'{e}zier curve we are using a Bernstein polynomial basis expansion where the control points are the coefficients or weights to the basis function. To fit the control points of a B\'{e}zier curve, consider the form of Equation (\ref{eq:bezier}). The structure of this equation is similar to a regression model. Using this as motivation, we can then naturally summarize the relationship between time points and control points with the following regression equation:
\begin{equation}
    \mathbf{y}_i = \mathbf{T}_i \bm{\theta} + \bm{\epsilon}_i
\end{equation}
The regression matrix, $\mathbf{T}_i$, has dimension $m_i \times (P + 1)$ and is evaluated at $t_{ij}$, where $t_{ij} \in [0, 1]$.

The fitting of B\'{e}zier curves can then be seen as equivalently fitting a multivariate linear regression.
\begingroup
\renewcommand*{\arraystretch}{1.5}
\begin{equation} \label{eq:reg_mat}
\mathbf{T}_i =
    \begin{bmatrix}
        b_0^P(t_{i1}) & b_1^P(t_{i1})  & \dots  & b_P^P(t_{i1})\\
        b_0^P(t_{i2}) & b_1^P(t_{i2})  & \dots  & b_P^P(t_{i2})\\
        \vdots & \vdots & \ddots & \vdots \\
         b_0^P(t_{i m_i}) & b_1^P(t_{i m_i})  & \dots  & b_P^P(t_{i m_i})
    \end{bmatrix}
\end{equation}
\endgroup

Now that we have the tools to fit a B\'{e}zier curve we will discuss how we use these curves to define cluster means and how to calculate how well an observed trajectory fits to a B\'{e}zier curve.

\section{Model Based Curve Clustering}

Assuming any route run by a player approximates one of the finite predefined routes, the goal of our work then becomes to try and classify each observed route as a realization of one of the existing predefined routes. This is equivalent to clustering the $n$ routes into $K$ clusters, which we will refer to as labeling the route. Labeling will be done iteratively with labels updated at each step. As such, we let $\mathbf{z} = (z_1, \ldots, z_n)$ be the current label of a given route, such that $z_i \in \{1, \ldots, K \}$. Our assumption claims that there exists some correct set of labels $\mathbf{z*}$ for our collection of routes. 

The regression matrix defined in (\ref{eq:reg_mat}) can be used to define the conditional Probability Density Function (PDF) of $\mathbf{y}_i$ given $\mathbf{t_i}$ as $f(\mathbf{y_i} | \mathbf{t_i}) = \mathcal{N}(\mathbf{y}_i | \mathbf{T}_i \bm{\theta}, \sigma^2 \mathbf{I})$.
Now we can consider a mixture of $K$ of these conditional distributions. 
Then the probability of observing the $i^\text{th}$ curve can be defined as:
\begin{equation}
    P(\mathbf{y}_i | \mathbf{T}_i, \bm{\theta}_k, \sigma^2_k) = \sum_{k = 1}^K \alpha_k \mathcal{N}(\mathbf{y}_i | \mathbf{T}_i\bm{\theta}_k, \sigma^2_k \mathbf{I}) 
\end{equation}
where $\alpha_{k}$ is the mixing weights of the $k^\text{th}$ cluster, and $\sum_{k=1}^K\alpha_{k}=1$. The log-likelihood of observing all $n$ curves is then defined as the log of the product of the probability of observing each curve. Which in turn is the sum over all the curves of the log of the probability of observing the curve: 
\begin{equation}
    \log( P( \mathbf{Y} | \mathbf{T}, \Theta) ) = \sum_{i = 1}^n \log \sum_{k = 1}^K \alpha_k \mathcal{N}(\mathbf{y}_i | \mathbf{T}_i\bm{\theta}_k, \sigma^2_k \mathbf{I}) 
\end{equation}
We have that $z_i$ is the cluster membership for curve $i$. Then the joint density of $\mathbf{y}_i$ and $z_i$ is
\begin{equation}
    P(\mathbf{y}_i, z_i | \mathbf{t}_i ) = \alpha_{z_i} \mathcal{N}(\mathbf{y}_i | \mathbf{T}_i\bm{\theta}_{z_i}, \sigma^2_{z_i} \mathbf{I}) 
\end{equation}
This leads us to use the Expectation Maximization (EM) Algorithm introduced by Dempster \cite{dempster} to learn the Maximum Likelihood Estimates (MLE) for a model with a latent variable. The algorithm consists of three steps: Initialization, Expectation, and Maximization.

We will first discuss here the general process for the Expectation and Maximization steps. We will discuss the specifics of the data pre-processing and initialization procedure in the next section.

\subsection{Expectation Step}

In this step, the current estimates are used to evaluate the conditional expectation. Based on Bayes’ rule, the membership probability $\pi_{ik}$ that the $i^\text{th}$ curve was generated from cluster $z_i$ is defined as
\begin{equation}
    \pi_{ik} = P(z_i = k | \mathbf{y}_i, \mathbf{T}_i)
\end{equation}
It can be calculated by computing the probability
that the $i^\text{th}$ curve is generated from cluster $k$
\begin{equation}
     \pi_{ik} = \alpha_k \frac{ P(\mathbf{y}_i | \mathbf{T}_i, z_i = k)}{P(\mathbf{y}_i | \mathbf{T}_i)} \\
\end{equation}
which is the product of the the probability of generating each $m_i$ observed points on the curve from cluster $k$.
\begin{equation}
    \pi_{ik} = \alpha_k  \frac{ \prod_{j=1}^{m_i} \mathcal{N}(\mathbf{y}_{ij} | \mathbf{T}_{ij}\bm{\theta}_{k}, \sigma^2_{k} \mathbf{I})}{\sum_{k = 1}^K \prod_{j=1}^{m_i} \mathcal{N}(\mathbf{y}_{ij} | \mathbf{T}_{ij}\bm{\theta}_{k}, \sigma^2_{k} \mathbf{I}) }
\end{equation}
In practice we can scale each $\mathcal{N}(\mathbf{y}_{ij} | \mathbf{T}_{ij}\bm{\theta}_{k}, \sigma^2_{k} \mathbf{I})$ by a constant without changing $\pi_{ik}$ to prevent underflow errors.

With regards to implementation, the heaviest computational part of calculating $\pi_{ik}$ is computing $\mathcal{N}(\mathbf{y}_{ij} | \mathbf{T}_{ij}\bm{\theta}_{k}, \sigma^2_{k} \mathbf{I})$ for every observed point. In practice, this calculation is performed in parallel in order to efficiently implement the algorithm.

\subsection{Maximization Step}

The updates rules for the parameters $\bm{\theta}_k, \sigma^2_k, \alpha_k$ are found by maximizing the log-likelihood. $\hat{\alpha}_k$ is the mean posterior probability that the $i^\text{th}$ curve was generated from cluster $z_i$, and $\pi_{ik}$ is obtained from the E-step:
\begin{equation}
    \hat{\alpha}_k = \frac{1}{n} \sum_{i = 1}^n \pi_{ik}
\end{equation}
$\hat{\bm{\theta}}_k, \hat{\sigma}^2_k$ are found through weighted least squares where the weights matrix $\mathbf{W}_k$ is a diagonal matrix with the diagonal elements the elements of the vector $\mathbf{w}_k$, defined as 
\begin{equation}
    \mathbf{w}_{k} = [ \underbrace{ 
    \underbrace{
        \pi_{1k},  \ldots, \pi_{1k}
      }_{\text{$m_1$~elements}},
      \underbrace{
        \pi_{2k},  \ldots, \pi_{2k}
      }_{\text{$m_2$~elements}},
      \ldots
      \underbrace{
        \pi_{nk},  \ldots, \pi_{nk}
      }_{\text{$m_n$~elements}}
      }_{\text{$\sum_{i=1}^n m_i$~elements}}]
\end{equation}
This leads to the following weighted least squares solutions (see for example \cite{gaffney}, \cite{chamroukhi}, \cite{faria})
\begin{equation}
    \hat{\bm{\theta}}_k = (\mathbf{T}^\intercal \mathbf{W}_k \mathbf{T} )^{-1} \mathbf{T}^\intercal \mathbf{W}_k \mathbf{Y}
\end{equation}
\begin{equation}
    \hat{\sigma}^2_k = \frac{1}{\sum_{i = 1}^n \pi_{ik}}( \mathbf{Y} - \mathbf{T}\hat{\bm{\theta}}_k )^\intercal \mathbf{W}_k ( \mathbf{Y} - \mathbf{T}\hat{\bm{\theta}}_k )
\end{equation}
Since we have assumed the variance covariance matrix of the Normal distribution is diagonal we take only the diagonal elements of $\hat{\sigma}^2_k$. 

In practice, the parameters $\hat{\bm{\theta}}_k$ and $\hat{\sigma}^2_k$ for different clusters can be computed in parallel and the $\mathbf{W}_k$ matrices can be stored as sparse matrices. This gives performance improvements when the number of clusters gets large.

We repeat the E and M steps until the change in log likelihood reaches a pre-defined tolerance ($10^{-6}$ in our case).

\section{Route Identification}
\subsection{NFL Tracking Data}

The data has a play ID variable that contains all players positional and directional data within each play. There is an event variable indicating the start, end, and key moment within each play. These events can be easily pulled from the data to identify the play ID for each passing play, which contains the corresponding position of each player on the field for the play's duration. We only looked at those events indicating a pass was attempted (6963 plays). Additionally only trajectories of offensive players who play Wide Receiver, Tight End, Running Back, or Full Back are analyzed (33967 trajectories). While other players can be eligible receivers we have no clear way of identifying them in the data. 

\subsection{Pre-processing}

We now perform 3 transformations to the trajectory data.

\begin{figure}[H]
    \centering
    \includegraphics[width=0.5\textwidth]{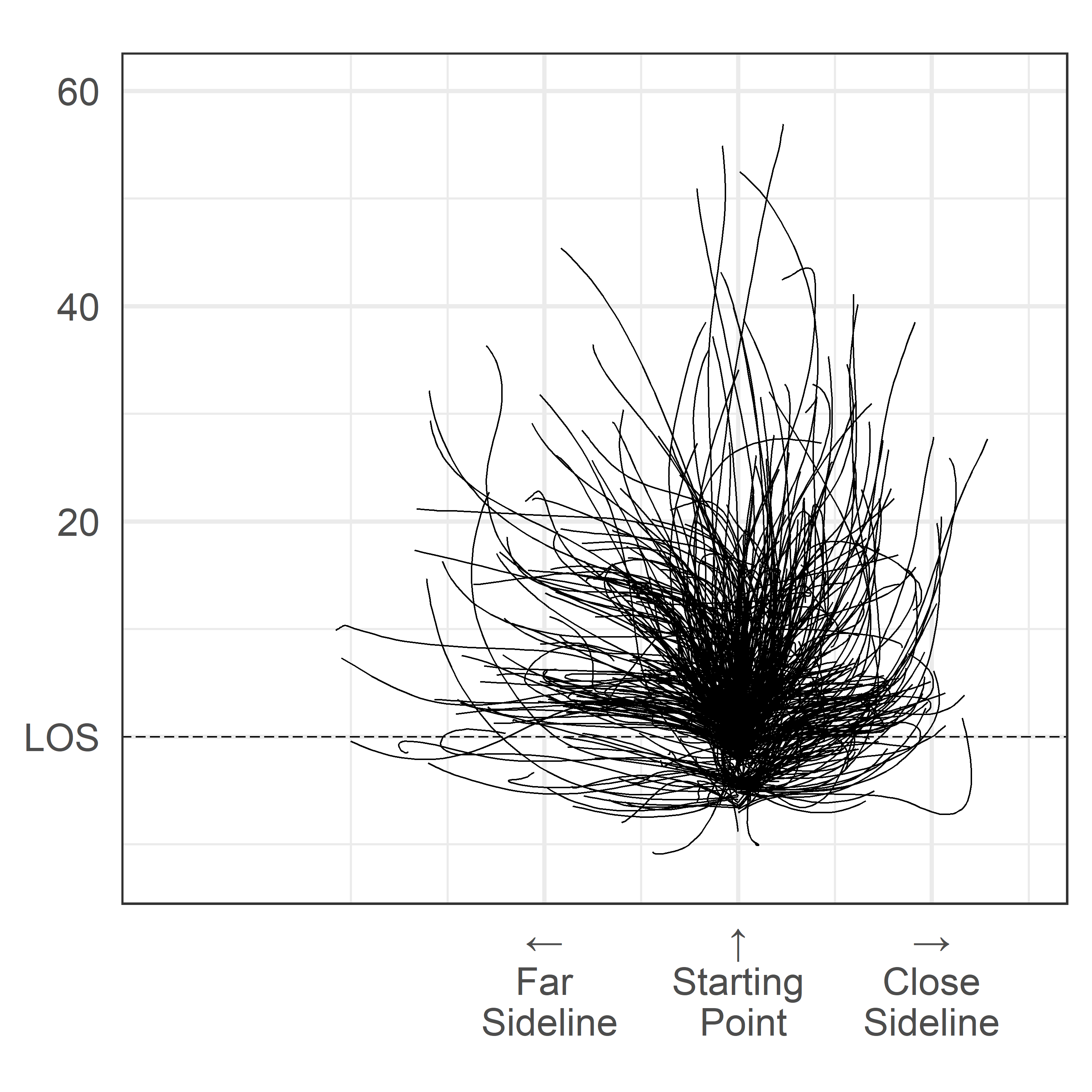}
    \caption{A sample of 500 transformed curves according to our pre-processing steps}
    \label{fig:pre-processed}
\end{figure}

\begin{enumerate}
    \item \textbf{Cut Trajectories}: Each player trajectory in its raw form starts before the event ``\textit{ball snap}'', and continues until the play is over, indicated by a number of possible events defined in the appendix. We only consider trajectory data beginning from when the ball is snapped, and ending when the play is over; the rest is cut from our analysis. The shift before the snap is outside the scope of this paper.

    \item \textbf{Standardize to Line of Scrimmage and Play Direction}: Since offensive plays start at varying distances from the target end zone, and the target end zone shifts for a given team every quarter; trajectories that represent similar routes will appear to be different x y space. Therefore we move all routes to a common line of scrimmage and orient player trajectories so that all routes are moving towards a common end zone.
    
    \item \textbf{Flip Trajectories}: We would expect any route run from one side of the quarterback to be approximately the mirror of the same route run from the other side (see Figure \ref{fig:route_tree}). All receiver routes are flipped to start from the same side of the quarterback, and translated to a single point of origin (see Figure \ref{fig:pre-processed})
\end{enumerate}
These pre-processing steps make it easier to identify common patterns and route labels. A shallow "in" route should look approximately the same now regardless of who ran it, and from where on the field they started from at the snap.

These simplifications make it easier to identify common patterns and route labels. An example of what the pre-processed data looks like is available in Figure \ref{fig:pre-processed}. After the clustering process we use the features of the non-transformed data for further investigation.

\subsection{Initialization}

Finally, in order to run the EM algorithm we need initial values for the curve centers. We determine these curve centers by assigning each observed curve to an initial cluster and then calculating initial control points, variance parameters, and cluster weights based on the curves in each cluster.

We use k-means clustering on the last observed point on each observed trajectory to use for the initialization of the cluster centers. This is a sensible idea since each trajectory has already been transformed to start at the same point, so much of the information about what route was run on the play is available in the last observed point. 

This is implemented on $K=30$ clusters. After getting our initial weights and parameters, the EM algorithm is run for four steps. Unfortunately the data does not include the true route run by each player, so labelling the curves was done manually.

\subsection{Implementation}

We implemented the curve clustering algorithm on the data for $K = 30$ clusters. This data consists of 33,967 routes from 6,963 passing plays. In total there are 1,438,133 measurements for an average of 42 measurements per route.

We used a computer with 8 CPUs and 52 GB of RAM. On average each Expectation step takes 1,910 seconds and each Maximization step takes 19 seconds. We run the algorithm for 4 steps. The log likelihood at each step is -6090879, -6077207, -6069274, -6064259. The total run time was 7,745 seconds. 
\begin{figure}[H]
    \centering
    \includegraphics[width=0.5\textwidth]{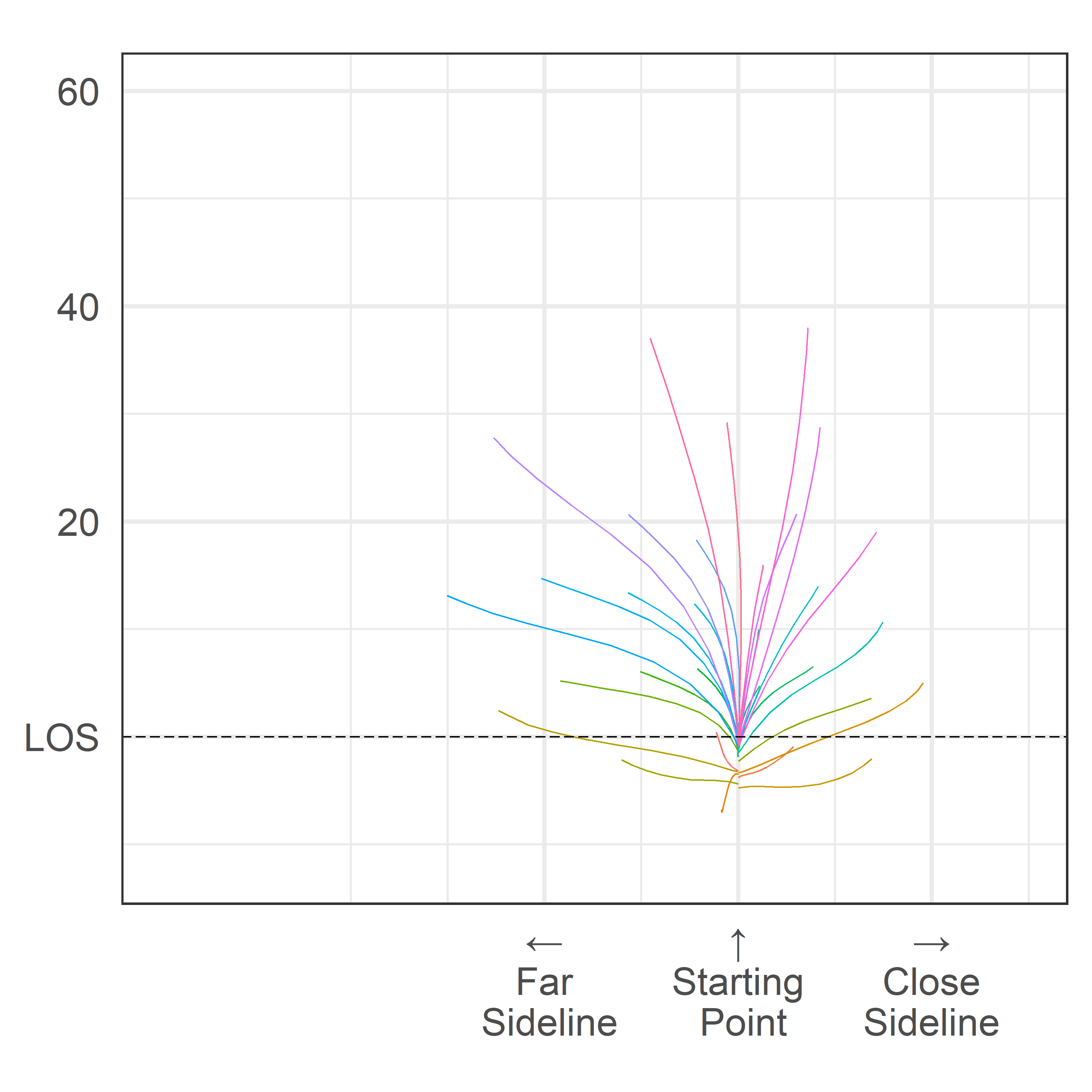}
    \caption{Cluster Means for 30 Clusters}
    \label{fig:clustered}
\end{figure}
\subsection{Route Labelling}

\begin{figure}[H]
    \centering
    \includegraphics[width=0.7\textwidth]{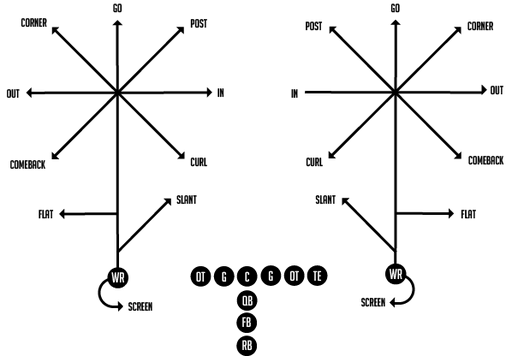}
    \caption{Example of a Route Tree \cite{route_tree}.}
    \label{fig:route_tree}
\end{figure}

\begin{figure}[H]
    \centering
    \includegraphics[width=\textwidth]{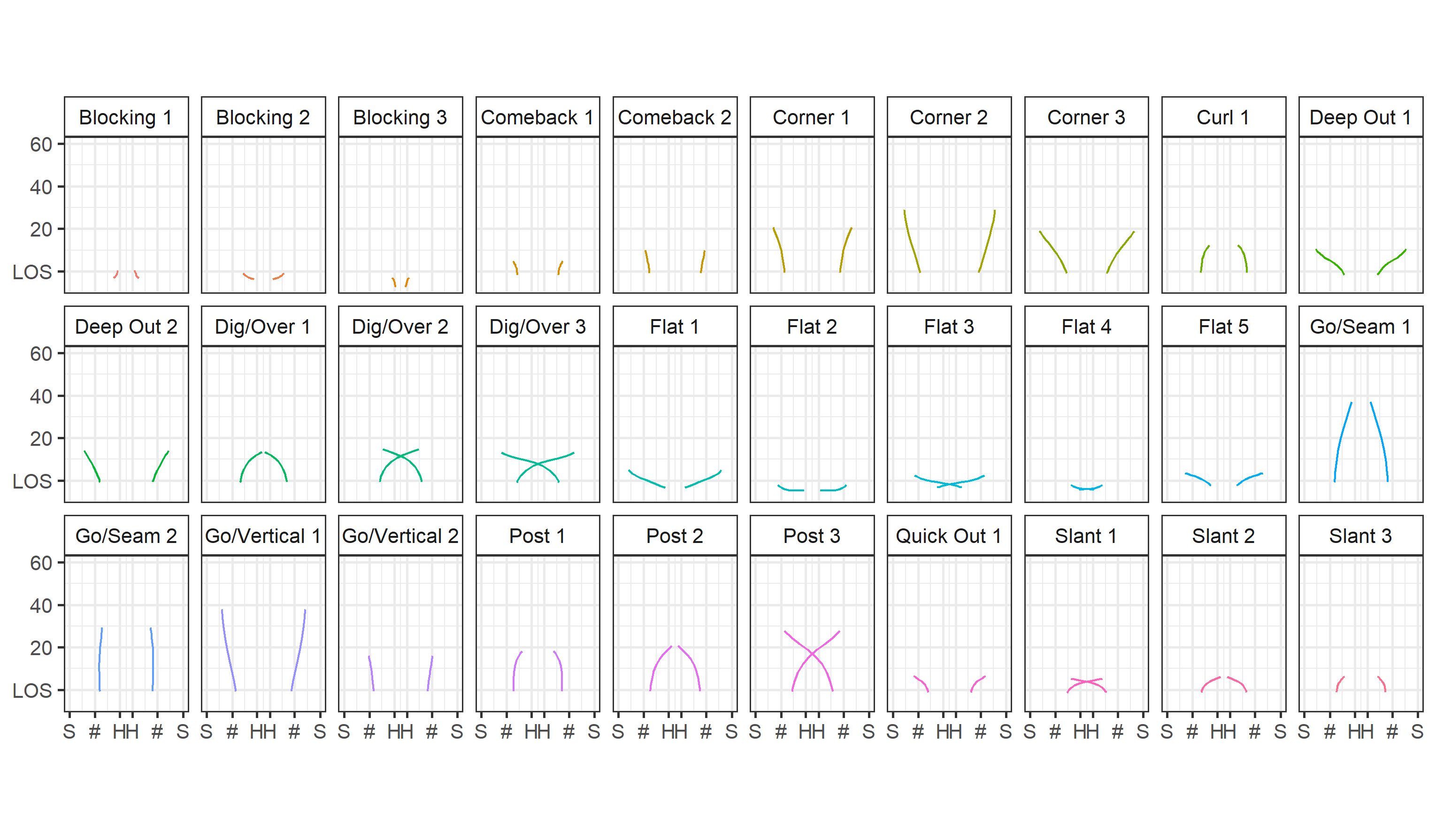}
    \caption{Labelled cluster means plotted with respect to the pre-transformed space}
    \label{fig:labelled}
\end{figure}

\begin{figure}[H]
    \centering
    \includegraphics[width=0.8\textwidth]{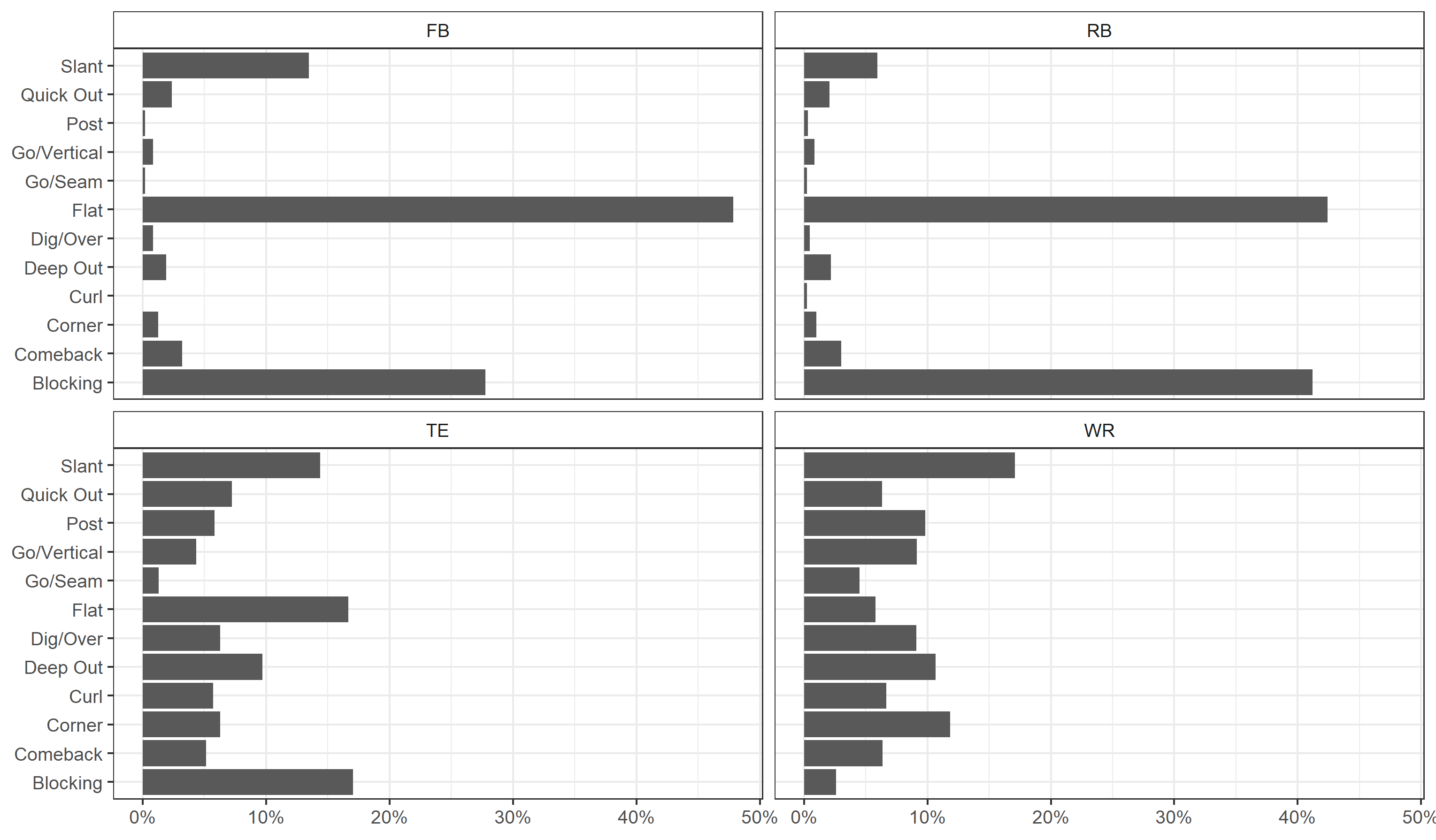}
    \caption{Routes per Position}
    \label{fig:position}
\end{figure}

\begin{figure}[H]
    \centering
    \includegraphics[width=\textwidth]{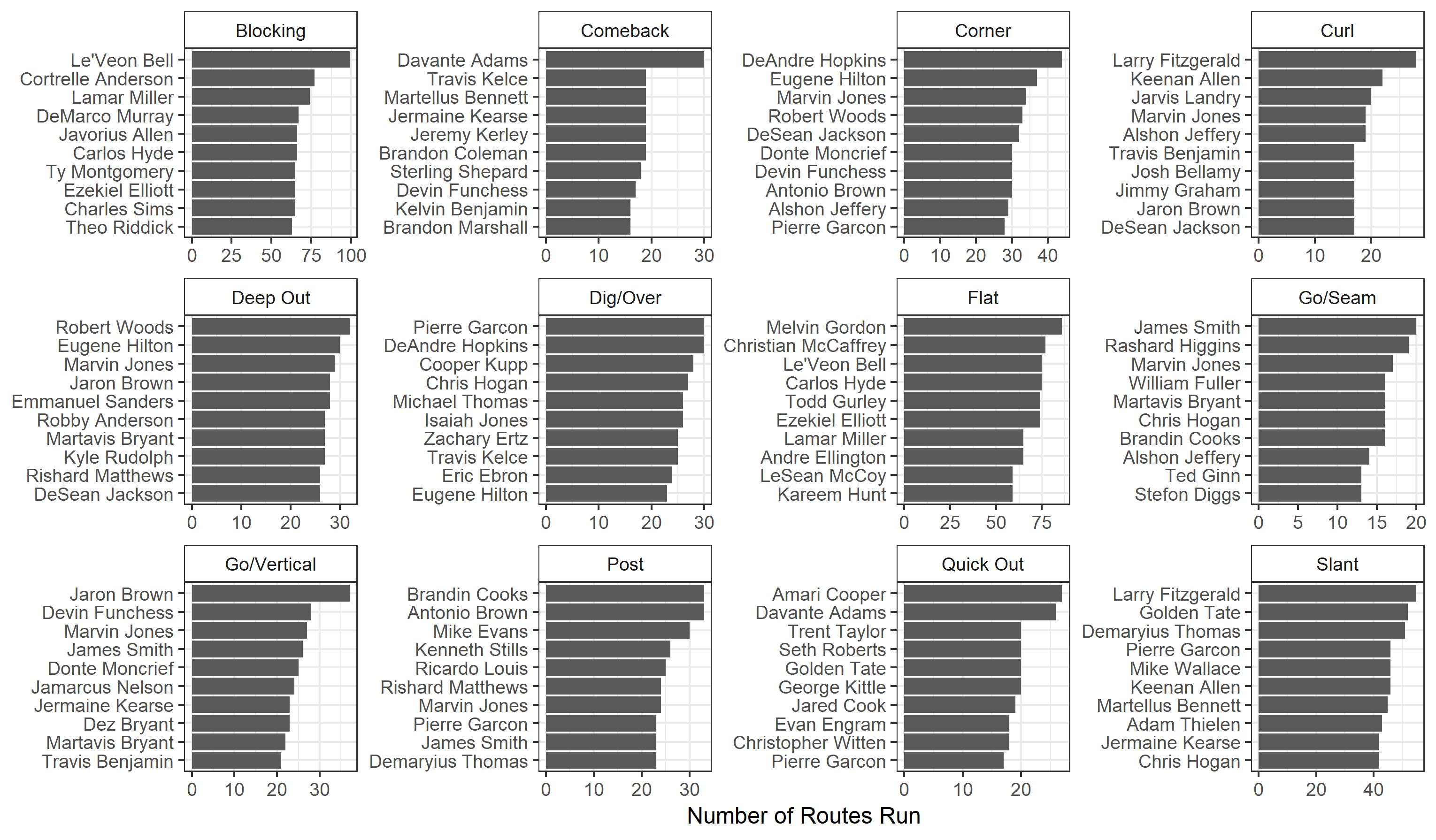}
    \caption{Players who run the most of each route}
    \label{fig:players-per-route}
\end{figure}

\begin{figure}[H]
    \centering
    \includegraphics[width=\textwidth]{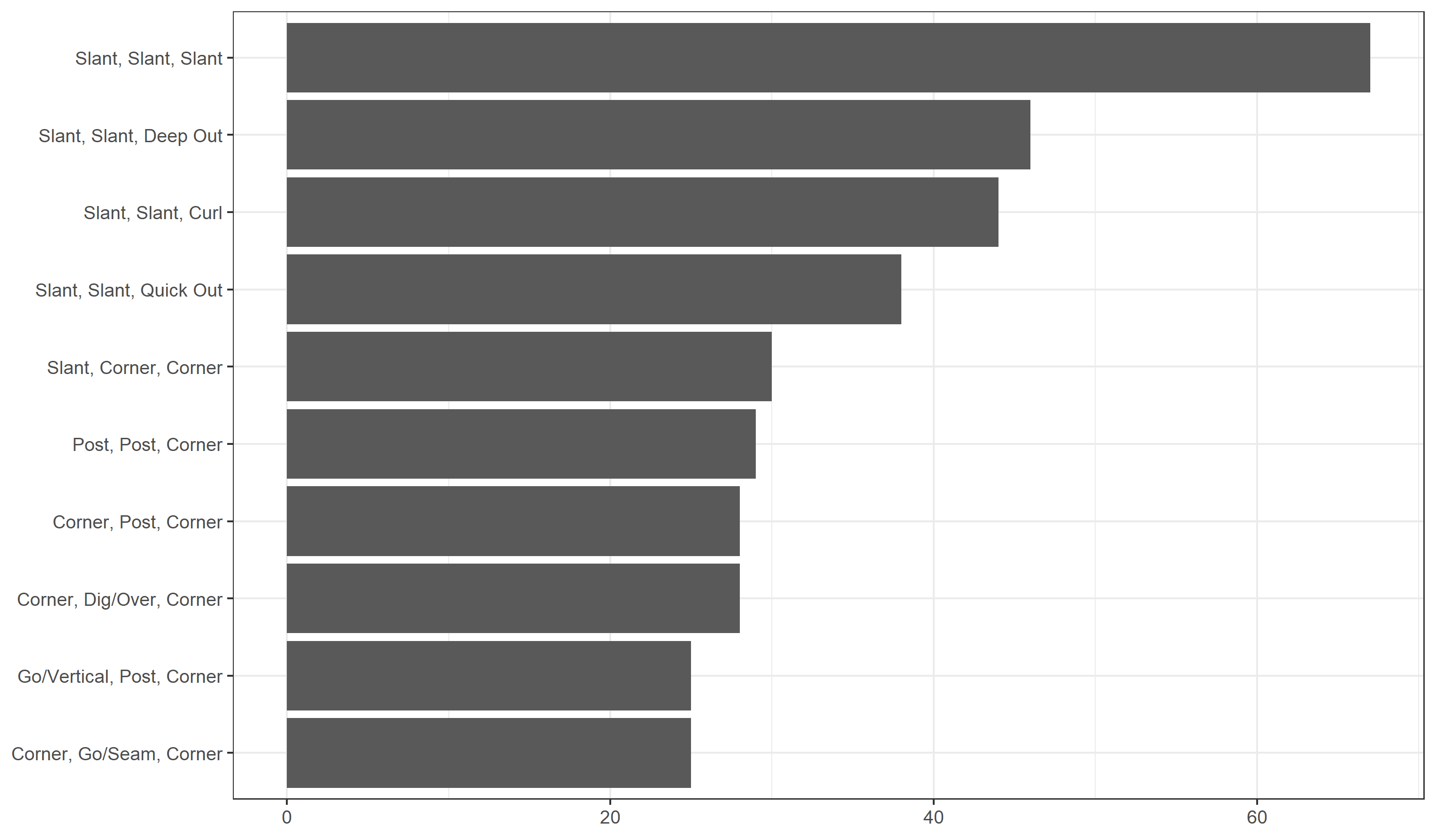}
    \caption{3 WR Designs}
    \label{fig:wr-designs}
\end{figure}

We have identified routes with similar structure but it is now our goal to add football context to our work by labelling the clusters. The cluster means obtained from our curve clustering process resemble those of route trees in football see Figure \ref{fig:route_tree}.
\begin{figure}[H]
  \centering
  \begin{minipage}[b]{0.49\textwidth}
    \includegraphics[width=\textwidth, height = 2.1 in]{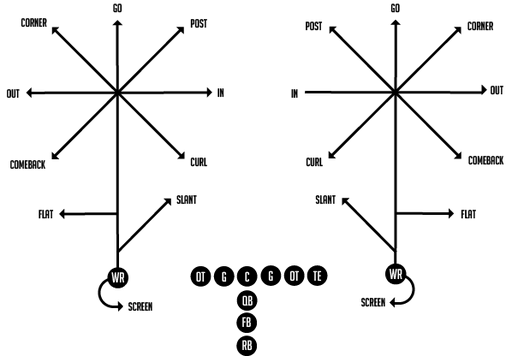}
    \subfloat{(a) Route tree for offensive receivers}
  \end{minipage}
  \hfill
  \begin{minipage}[b]{0.49\textwidth}
    \includegraphics[width=\textwidth]{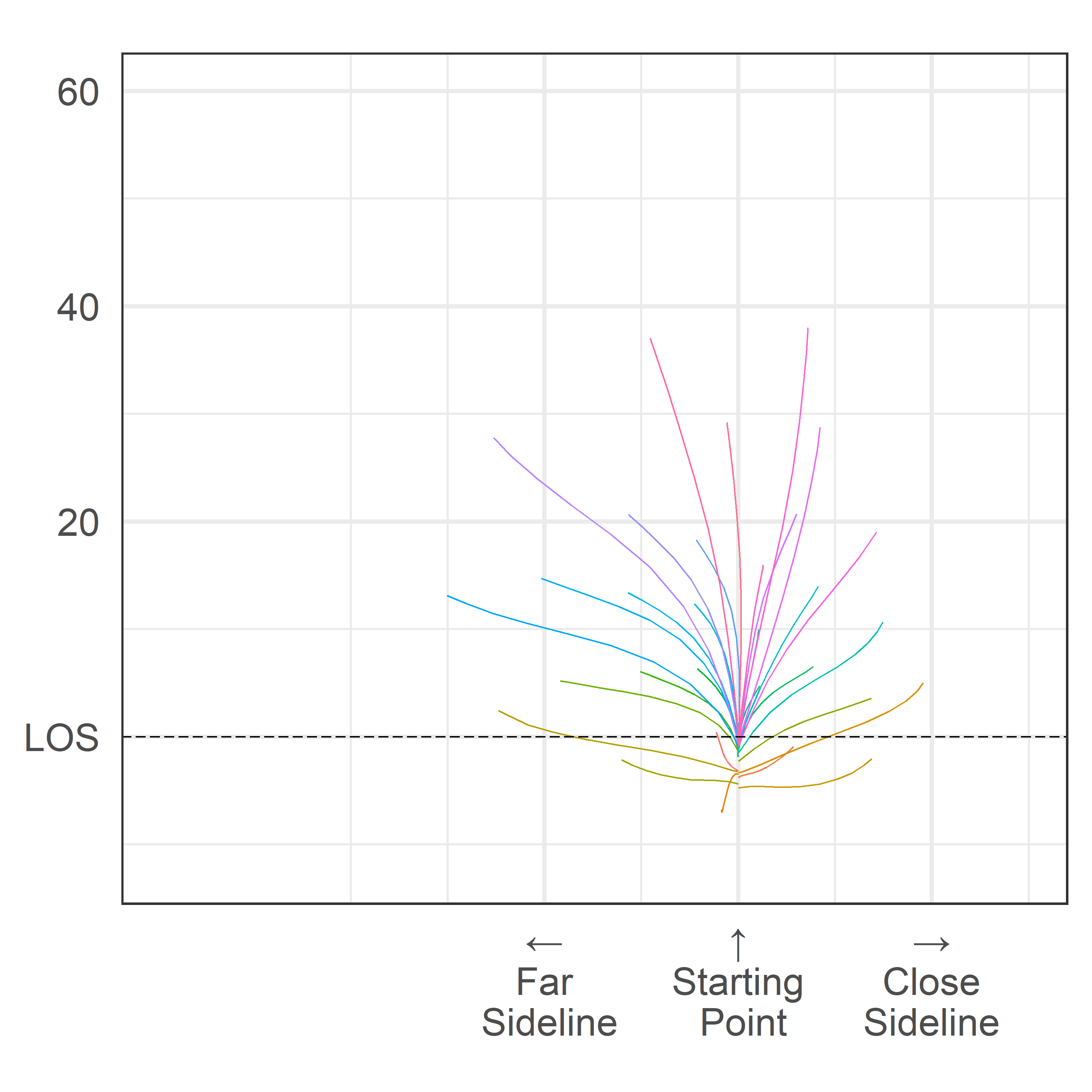}
    \subfloat{(b) Clustered means}
  \end{minipage}
  \caption{Route tree (a) when compared to the results of the clustered means (b)}
\end{figure}
We use labels provided by Ben Minaker formerly of the Simon Fraser University Football Team to manually label the mean curve of each cluster. Some clusters end up representing similar routes so we end up condensing the 30 clusters into 12 route groups. These route groups can be plotted back in ``football space'' and are displayed in Figure
\ref{fig:labelled}.
This then leads us to perform brief sanity checks that our labelling process worked. In Figure \ref{fig:position} we can compare the route distribution across positions. Despite, the algorithm having limited indicators of a player's position the routes assigned to each position align with our intuition of player positions.
Finally, we can look at very basic trends about route usage. For example we could look at Figure \ref{fig:players-per-route} to see which players run which routes most often or Figure \ref{fig:wr-designs} to see which 3 WR design play concepts are run most often.
These plots provide basic information provided by our labelled routes. However, we believe that there is much more that can be discovered by using route labels in analysis. We have implemented basic versions of them but leave it to future work to develop them fully.

\section{Future Work}

There are 2 main directions for future work in this space. The first is to improve the clustering process. One way in which this can be achieved is by improving the computational efficiency of the clustering process. This would allow for more clusters to be fit. We've considered that perhaps clustering based on derivatives and second derivatives of the position vectors may yield improved results. A major drawback is the ability to identify a comeback vs. a go route as the functions look nearly identical to our clustering algorithm. Derivatives of the function will show velocity over time, and second derivatives acceleration. Finally, there are more complex models in \cite{gaffney} that may yield better clustering results.
 
The second is to augment the analysis of football players. As mentioned in the previous section, we see major opportunities to use these labelled routes to understand and account for player usage across receiver statistics.

Additionally, the results obtained thus far cannot be proven to work without knowing the true route name for each of the passing plays in our data set. Our only method for checking the results is to compare to our intuition of player tendencies. In the future we hope to work with teams to calibrate our algorithm for accuracy route labelling. 

\subsection{Potential Uses}

The automated labelling of routes provides potential avenues for more in depth analysis, e.g. with labelled routes on each play we can build models to understand player deployment, receiver statistics that account for usage \cite{tae}, and build better defensive statistics for coverage players. 

\begin{figure}[H]
    \centering
    \includegraphics[width=\textwidth]{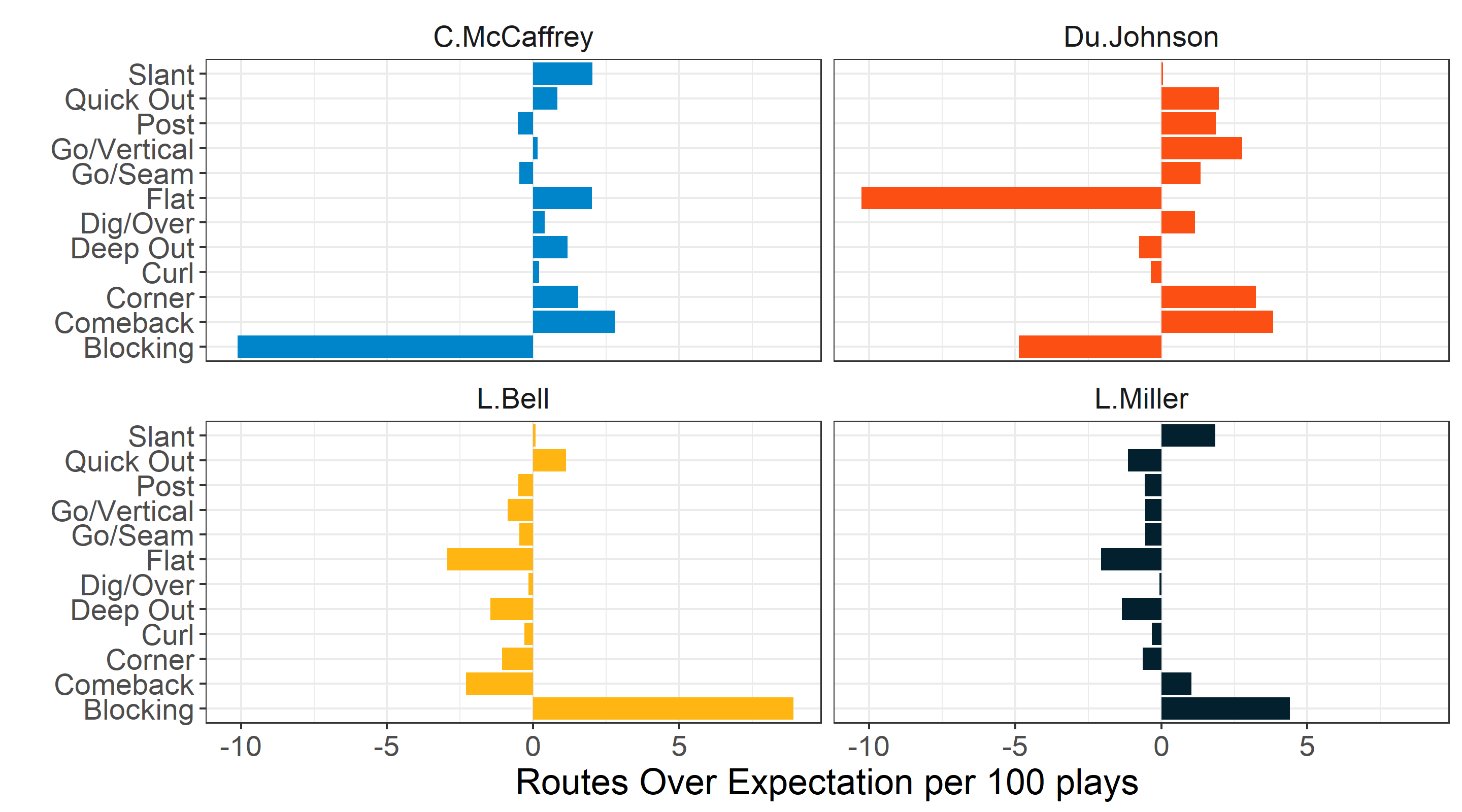}
    \caption{RB Routes Run Over Expectation per 100 Plays}
    \label{fig:rb-usages}
\end{figure}

\begin{figure}[H]
    \centering
    \includegraphics[width=\textwidth]{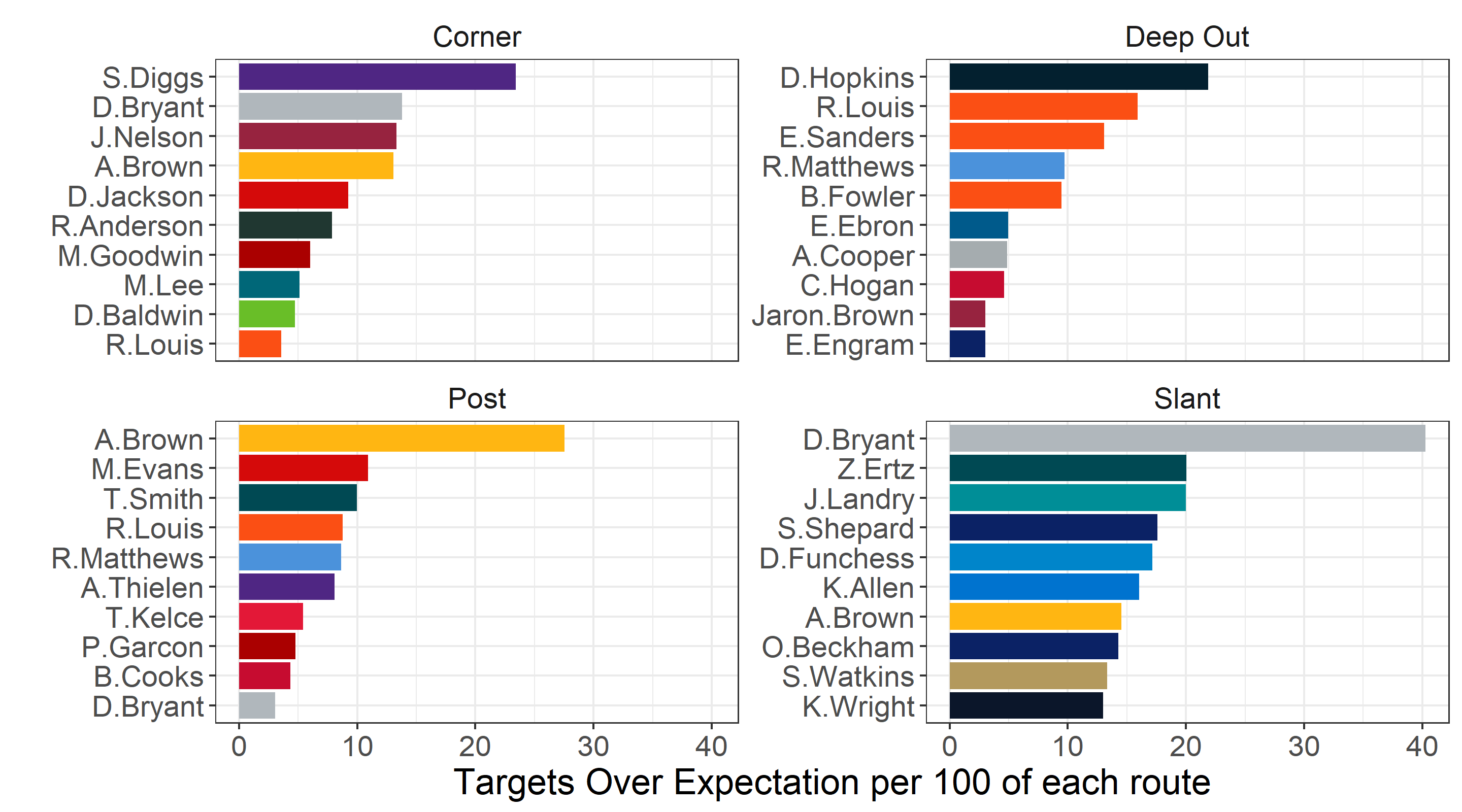}
    \caption{Targets Over Expectation per 100 of each Route by Route}
    \label{fig:tae-route}
\end{figure}

\begin{figure}[H]
    \centering
    \includegraphics[width=\textwidth]{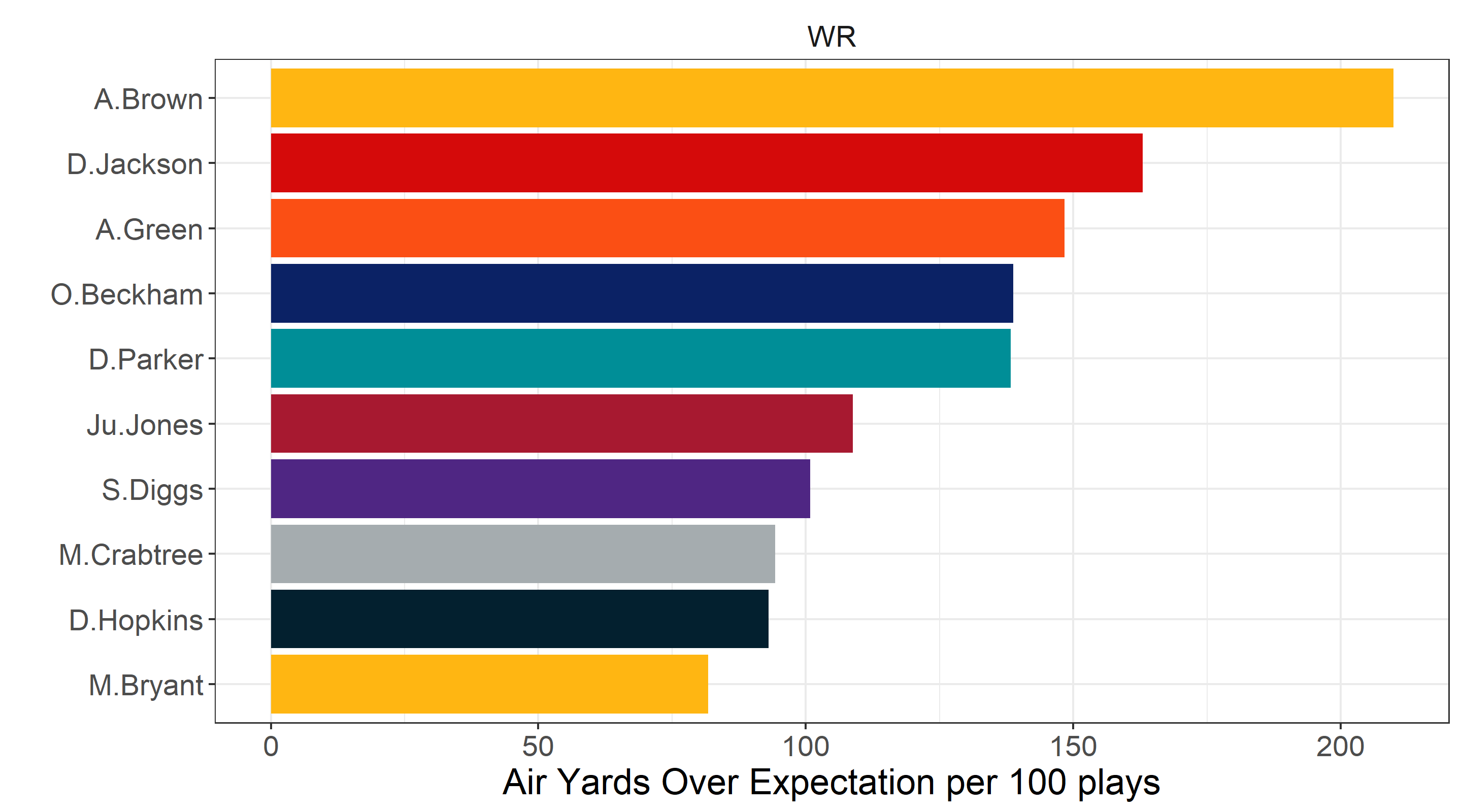}
    \caption{Air Yards Over Expectation per 100 Routes}
    \label{fig:ayoe-player}
\end{figure}

We can calculate statistics like type of route run over expectation per 100 plays (see Figure \ref{fig:rb-usages}) to understand player usage compared to an average player while accounting for position and game situation. These can then be further used to cluster players based on their usages above expectation of an average player.
This can then be extended to replicate the work of \cite{tae} to compute targets over expectation for various wide receivers. This can be broken down in a number of different ways. In Figure \ref{fig:tae-route} we break down Targets Over Expectation by 4 different routes to see which receivers are being targeted more often then expected on specific route types.
This methodology can be extended to the concept of Air Yards. As provided in \cite{R-nflscrapR}, Air Yards can be used to contextualize and quantify receiving opportunities. We plot in Figure \ref{fig:ayoe-player} the top players for Air Yards Over Expectation per 100 Routes from a preliminary model.
As we have demonstrated briefly here there is a large array of applications that can use route labels. One that we did not provide an example for is improving the evaluation of defensive players. The labelling of routes from tracking data is therefore valuable for further analysis as it is for automated film tagging.

\section{Conclusion}

In this work we demonstrate a method for labelling routes for player trajectory data of varying lengths in the National Football League. We do so by using a model based clustering approach and the Expectation Maximization algorithm. The probabilistic model works as a mixture of Gaussian distributions centered at B\'{e}zier curves. This provides the potential to understand player usages, efficiencies, and improve defensive evaluation.



\section{Acknowledgements}
  
We would like to thank Coco Liu, Barinder Thind and Cherlene Lin for their helpful discussions around the EM algoirthm and functional data analysis, Michael Couture and Ben Minaker for their help with football interpretations and Tim Swartz for his guidance and supervision. To the developers of the tidyverse family of packages \cite{R-tidyverse} without which we would not have been able to perform this research. We would also like to thank the football analytics community on twitter for their support and guidance throughout this project. Finally, we wanted to thank Michael Lopez and the NFL for hosting this competition and giving us this opportunity.

\bibliographystyle{apalike}
\bibliography{references}

\begin{thebibliography}{}

\bibitem[Aghabozorgi et~al., 2015]{aghabozorgi}
Aghabozorgi, S., Shirkhorshidi, A.~S., and Wah, T.~Y. (2015).
\newblock Time-series clustering – a decade review.
\newblock {\em Information Systems}, 53:16--38.

\bibitem[Ajmeri and Shah, 2012]{ajmeri}
Ajmeri, O. and Shah, A. (2012).
\newblock Using computer vision and machine learning to automatically classify
  nfl game film and develop a player tracking system.
\newblock In {\em Proceedings of the 2012 MIT Sloan Sports Analytics
  Conference}.

\bibitem[AlShaher, 2018]{alshaher}
AlShaher, A.~A. (2018).
\newblock Arabic character recognition using regression curves with the
  expectation maximization algorithm.
\newblock {\em International Journal of Computer, Electrical, Automation,
  Control and Information Engineering}, 12(12):1087--1091.

\bibitem[Baumer et~al., 2013]{open_war_b}
Baumer, B., Jensen, S., and Matthews, G. (2013).
\newblock Openwar: An open source system for evaluating overall player
  performance in major league baseball.
\newblock {\em Journal of Quantitative Analysis in Sports}, 11.

\bibitem[Bouveyron and Jacques, 2011]{Bouveyron2011}
Bouveyron, C. and Jacques, J. (2011).
\newblock Model-based clustering of time series in group-specific functional
  subspaces.
\newblock {\em Advances in Data Analysis and Classification}, 5(4):281--300.

\bibitem[Broadie, 2011]{broadie}
Broadie, M. (2011).
\newblock Assessing golfer performance on the pga tour.
\newblock {\em Interfaces}, 42.

\bibitem[Buccaneers.com, 2015]{route_tree}
Buccaneers.com (2015).
\newblock Red chalk talk: Route tree (3 of 4).
\newblock [Online; posted 30-August-2015].

\bibitem[{Chamroukhi}, 2013]{chamroukhi}
{Chamroukhi}, F. (2013).
\newblock Robust em algorithm for model-based curve clustering.
\newblock {\em arXiv e-prints}, page arXiv:1312.7022.

\bibitem[Dempster et~al., 1977]{dempster}
Dempster, A.~P., Laird, N.~M., and Rubin, D.~B. (1977).
\newblock Maximum likelihood from incomplete data via the em algorithm.
\newblock {\em Journal of the Royal Statistical Society. Series B
  (Methodological)}, 39(1):1--38.

\bibitem[Dong et~al., 2018]{jiguo}
Dong, J.~J., Wang, L., Gill, J., and Cao, J. (2018).
\newblock {{F}unctional principal component analysis of glomerular filtration
  rate curves after kidney transplant}.
\newblock {\em Stat Methods Med Res}, 27(12):3785--3796.

\bibitem[Faria and Soromenho, 2010]{faria}
Faria, S. and Soromenho, G. (2010).
\newblock Fitting mixtures of linear regressions.
\newblock {\em Journal of Statistical Computation and Simulation},
  80(2):201--225.

\bibitem[Gaffney, 2004]{gaffney}
Gaffney, S. (2004).
\newblock {\em Probabilistic Curve-Aligned Clustering and Prediction with
  Mixture Models}.
\newblock PhD thesis, University of California, Irvine.

\bibitem[Hochstedler and Gagnon, 2017]{hochstedler}
Hochstedler, J. and Gagnon, P.~T. (2017).
\newblock American football route identification using supervised machine
  learning.
\newblock In {\em Proceedings of the 2017 MIT Sloan Sports Analytics
  Conference}.

\bibitem[Horowitz et~al., 2018]{R-nflscrapR}
Horowitz, M., Yurko, R., and Ventura, S. (2018).
\newblock {\em nflscrapR: Compiling the NFL Play-by-Play API for easy use in
  R}.
\newblock R package version 1.8.1.

\bibitem[Leroy et~al., 2018]{Leroy_2018}
Leroy, A., MARC, A., DUPAS, O., REY, J.~L., and Gey, S. (2018).
\newblock Functional data analysis in sport science: Example of swimmers'
  progression curves clustering.
\newblock {\em Applied Sciences}, 8(10):1766.

\bibitem[McNicholas and Murphy, 2010]{mcnicholas}
McNicholas, P.~D. and Murphy, T.~B. (2010).
\newblock Model-based clustering of microarray expression data via latent
  gaussian mixture models.
\newblock {\em Bioinformatics}, 26(21):2705--2712.

\bibitem[Miller and Bornn, 2017]{Miller2017Possessions}
Miller, A.~C. and Bornn, L. (2017).
\newblock Possession sketches : Mapping nba strategies.
\newblock In {\em Proceedings of the 2017 MIT Sloan Sports Analytics
  Conference}.

\bibitem[N.~Bernstein, 1911]{bernstein}
N.~Bernstein, S. (1911).
\newblock D\'{e}monstration du th\'{e}or\`{e}me de weierstrass fond\'{e}e sur
  le calcul des probabilit\'{e}s.
\newblock {\em Communications de la Soci\'{e}t\'{e} Math\'{e}matique de Kharkov
  2}, 13.

\bibitem[Nba, 2013]{nba}
Nba (2013).
\newblock Nba partners with stats llc for tracking technology.
\newblock [Online; posted Sep 5, 2013].

\bibitem[Nfl, 2019]{nfl_tracking}
Nfl (2019).
\newblock Nfl next gen stats.
\newblock https://operations.nfl.com/the-game/technology/nfl-next-gen-stats/.
\newblock Accessed: 2019-04-23.

\bibitem[Rossler, 2019]{tae}
Rossler, B. (2019).
\newblock Introducing targets above expectation.

\bibitem[Stern, 1994]{stern}
Stern, H.~S. (1994).
\newblock A brownian motion model for the progress of sports scores.
\newblock {\em Journal of the American Statistical Association},
  89(427):1128--1134.

\bibitem[Wickham, 2017]{R-tidyverse}
Wickham, H. (2017).
\newblock {\em tidyverse: Easily Install and Load the 'Tidyverse'}.
\newblock R package version 1.2.1.

\bibitem[Yurko et~al., 2019]{nflwar}
Yurko, R., Ventura, S., and Horowitz, M. (2019).
\newblock nflwar: a reproducible method for offensive player evaluation in
  football.
\newblock {\em Journal of Quantitative Analysis in Sports}.

\end{thebibliography}

\end{document}